\documentclass[journal=jacsat,manuscript=article]{achemso}

\usepackage{chemformula} 
\usepackage[T1]{fontenc} 

\usepackage{hyperref}
\hypersetup{
    colorlinks=true,
    linkcolor=blue,
    filecolor=magenta,      
    urlcolor=gray,
    }

\mciteErrorOnUnknownfalse
\author{S. Saini}
\email{sainishubh4697@gmail.com}
\author{Ashok P}
\author{Amit Verma}
\email{amitkver@iitk.ac.in}
\affiliation[Unknown University]
{Department of Electrical Engineering, Indian Institute of Technology Kanpur, Kanpur, 208016, U.P., India}

\title[An \textsf{achemso} demo]
  {Dynamic Multi Color Switching using Ultrathin Vanadium Oxide on Aluminium based Asymmetric Fabry-P\'{e}rot Resonant Structure}

\begin{document}


\begin{abstract}
Vanadium dioxide ($VO_{2}$) exhibits strong infrared optical switching due to its insulator-metal phase-transition property. However, in the visible wavelengths, it's intrinsic optical switching is quite low. Current research explores solutions like multilayering, intricate structural patterning, high thermal budget processes and costly metals for improved color switching. Nonetheless, the color gamut coverage with these methodologies remains notably limited. This work overcomes these limitations and demonstrates dynamic multi-colour switching covering a large color gamut using a simple, unpatterned, ultrathin ($\sim$ $\frac{\lambda}{14}$, where wavelength $\lambda$ is taken as 575 nm at the center of visible spectrum) asymmetric Fabry-P\'{e}rot structure of $VO_{2}$ on Aluminium (Al). We use the transfer matrix method to design the $VO_{2}/Aluminium\,(Al)/Sapphire$ structure for maximum visible reflectance switching. $VO_{2}$ films are synthesized using a simple, low thermal budget atmospheric oxidation of Vanadium (V). With varying oxidation durations, different colors of the oxidized samples are observed. Consistent and reversible color-switching is observed visibly and in reflectance measurements with the change in temperature from low (RT $\sim$ 30$^{\circ}$C) to high (HT $\sim$ 100$^{\circ}$C) or vice versa due to the phase transition property of the $VO_{2}$ layer in the structure. Compared to the existing studies, this work shows a significant change in chromaticities and covers a large color gamut when plotted on the CIE chromaticity diagram. This work has potential applications in the fields of display, thermochromic structures, and visible camouflage.
\end{abstract}

\section{Introduction}\label{Introduction}
Dynamic/active color switching has received extensive research attention for its applications in smart glasses \cite{doi:10.1146/annurev-chembioeng-080615-034647, ke2016two}, optical filters \cite{kepic2021optically, he2020dynamically}, and display technology \cite{duan2017dynamic}. Typically, active color switching has been achieved by liquid crystals, nanoscale plasmonic devices, and phase-change/phase-transition materials (PCM/PTM) \cite{garshasbi2019using}. $VO_{2}$ is one such PTM, which changes its crystal structure from monoclinic to rutile as it switches from insulating to metallic state at $\sim 68^{\circ}C$ \cite{chain1991optical}. This reversible transition can be induced electrically \cite{wu2011electric}, thermally \cite{kawakubo1964phase}, optically \cite{liu2012terahertz}, or through strain \cite{cao2009strain}.\\
$VO_{2}$ has shown significant optical switching in the infrared region ($IR$) for different substrates like sapphire \cite{de1989formation, jin1997dependence}, glass \cite{ashok2020high}, ITO \cite{ashok2020high}, silicon \cite{saitzek2008vo2}, etc. However, its optical switching is very low in the visible region ($\sim$ 400 to 750 nm) of the electromagnetic spectrum \cite{graf2017interdependence, zhang2010thermochromic, liu2017dual}. Various studies report subduing this limitation using multilayering, nanoscale patterning, doping, etc \cite{khaled2021current}. For example in the simulation study by Song et al. \cite{song2018tailoring}, the multilayer structure of Ag nanoparticles, a thin layer of $VO_{2}$, and a thick Ag back mirror is reported for the dynamic color display. They predicted various colors like cyan, green, magenta, pink, blue, and aquamarine. Similarly, simulation work by K Wilson et al. \cite{wilson2018dynamic} shows multiple color changes by planar $ITO-VO_{2}-ITO-Ag$ thin film stacks. Shu et al. \cite{shu2018dynamic} proposed a structure having Ag nanodisks on $VO_{2}$ thin film; different colors were predicted by tuning the spatial periodicity and diameter of the Ag nanodisks. An experimental study by Liwei Zhou et al. \cite{zhou2018modified} uses Au nanoparticles \& $VO_{2}$ sandwiched structure ($VO_{2}/Au/VO_{2}$) to demonstrate change in optical response for smart windows application. Sungjun In et al. \cite{in2020self} demonstrated the thermochromic behavior with self-organized gold network-$VO_{2}$ (SGN-$VO_{2}$) hybrid film ($VO_{2}/Au$). The deposited gold film is annealed to undergo a change in morphology to form SGN-Au. These changes cause morphological strain and changes in grain sizes in the $VO_{2}$ layer. Depending on the thickness of $VO_{2}$ and Au, multiple colors like aquamarine, yellow, cyan, purple, light-violet, magenta, etc., were reported. Similarly, Zhao J. et al. \cite{zhao2021flexible} report vivid structural colors with temperature generated by multilayer Ag/$VO_{2}$/Al structure on rigid and flexible substrates. The thickness of $VO_{2}$ and Ag layers was changed to control the color hue, saturation and brightness of the structure. \\
Most of the existing studies use multilayer structures, nano-scale features, or costly metals like Ag and Au in various forms to achieve better visible color switching with $VO_{2}$. However, opting for these approaches significantly adds to the process cost and complexity. In some cases, precise control of process parameters and high thermal budget annealings are involved, adding further limitations to these approaches. In this work, we design and fabricate a uniform $VO_{2}$/Al/sapphire structure for active color switching. Deposited Vanadium (V) thin film is oxidized in atmospheric conditions to synthesize $VO_{2}$ thin films. Various colors are noticed during the oxidation process, and depending on the sample’s oxidation duration, different extent of color switching is obtained. Raman and reflectance measurements are used for the sample characterization. Samples containing $VO_{2}$ are found to exhibit strong and reversible color switching as a function of temperature. Table \ref{conclusion_comparision_table} provides a synopsis of the comparison between this work and the aforementioned studies; the specifics of this work are presented in the following text.

\section{Design and simulation}\label{Design and simulation}
\begin{figure}[H]
\centerline{\includegraphics[scale=0.6]{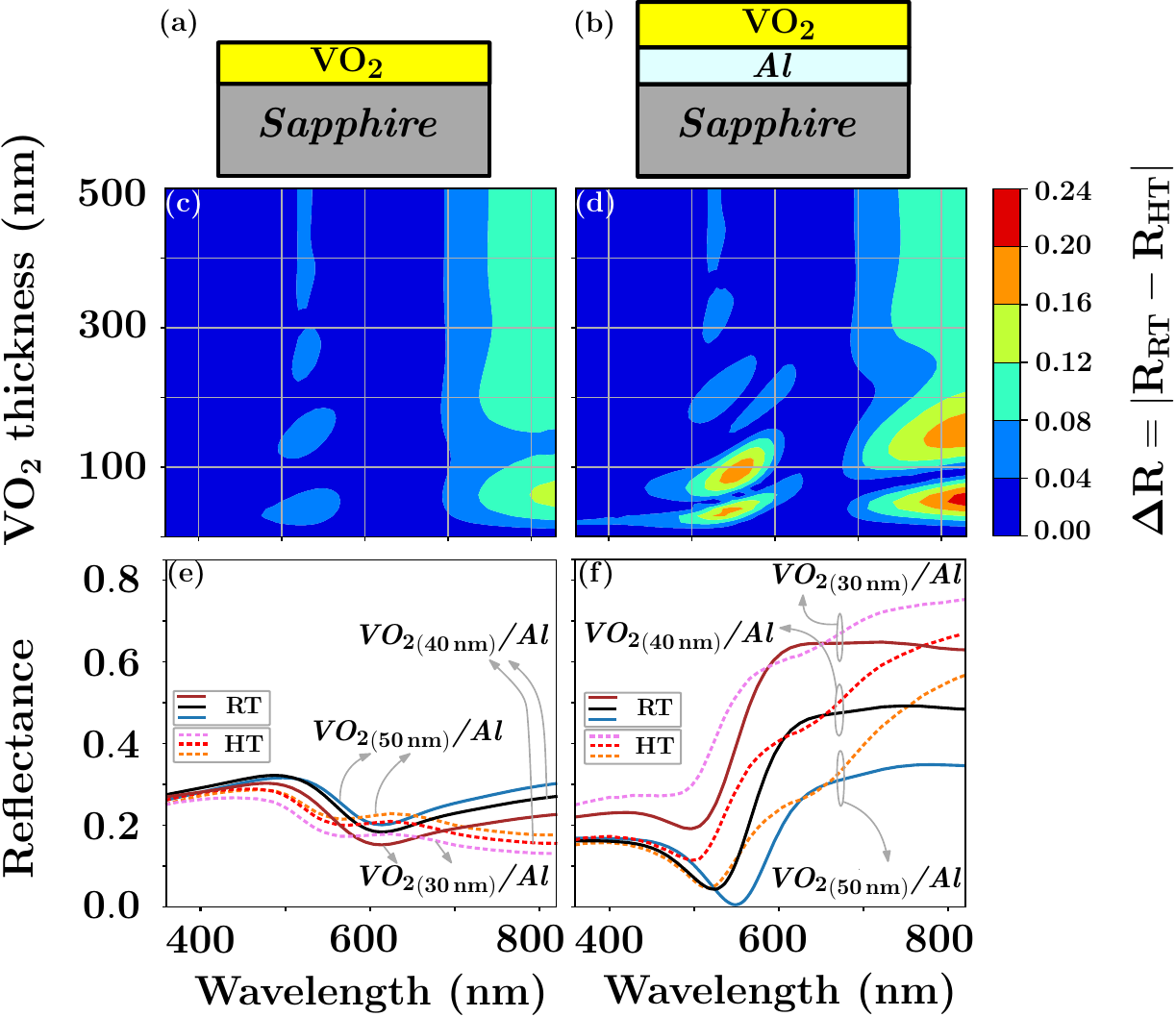}}
\caption{Schematic of the (a) $VO_{2}$/Sapphire and (b) $VO_{2}$/Al/Sapphire thermochromic structure. (c), (d) Simulated reflectance switching ($\Delta R$) at different wavelengths with varying $VO_{2}$ thicknesses for (a) and (b), respectively. Two hotspots of $\Delta R$ are observed for $VO_{2}$ thickness of 40 and 95 nm for (b). (e), (f) Comparison of calculated reflectance spectra at RT and HT for 30 nm, 40 nm and 50 nm of $VO_{2}$ on Sapphire (S) and Al/Sapphire (S), respectively. Compared to $VO_{2}/S$, the $VO_{2}/Al/S$ shows a significant change in calculated reflectance and spectral shift in minima with changes in thickness and temperature. These observed changes can be primarily attributed to the Fabry–P\'{e}rot resonances or interference effects in the $VO_{2}/Al/S$, akin to those observed in the conventional ultrathin dielectric-metal structures \cite{zhao2021flexible, kats2013nanometre}. The refractive indices of $VO_{2}$ are taken from \cite{wan2019optical}.}
\label{simulation}
\end{figure}

Standalone $VO_{2}$ suffers from low optical switching in the visible region. To overcome this, we propose having Al as a back reflector, eliminating the transmission loss. The reflected signals from the Al top surface would lead to constructive and destructive interference effects. To find the optimum thickness of Al, the transfer matrix method (TMM) simulations are performed. Python TMM library is used for the simulations \cite{pythontmm}. The results show zero transmission loss after 50 nm of Al thickness; further details can be found in the supplementary material. We analyze the reflectance in the insulating and metallic phase of $VO_{2}$ with varying thicknesses for the $VO_{2}$/Sapphire (S) and $VO_{2}$/Al/Sapphire (S) structures using the TMM. As mentioned earlier, $VO_{2}$ is in the insulating phase at room temperature (RT) and transitions to the metallic phase at high temperature (HT) beyond $68^{\circ}C$. These phases have different refractive indices \cite{wan2019optical}. The absolute difference between insulating and metallic phase reflectance (Reflectance switching $\Delta R = \lvert R_{RT} - R_{HT} \rvert$) is used to quantify the extent of optical switching. The two simulated structures, $VO_{2}$/S and $VO_{2}$/Al/S are shown in Fig. \ref{simulation} (a) and \ref{simulation} (b), respectively. Fig. \ref{simulation} (c) and \ref{simulation} (d) show the contour plots of the $\Delta R$ with respect to the wavelength and varying thicknesses of $VO_{2}$ for the structures shown in Fig. \ref{simulation}(a) and \ref{simulation} (b), respectively.\\
On comparing the results, it can be observed that $VO_{2}$/Al/S shows enhanced optical switching for smaller thickness ($\leq$ 200 nm) than $VO_{2}$/S; the results also show two hotspots of $\Delta R$ near $\sim$ 95 nm and 40 nm of $VO_{2}$. This is due to Fabry–P\'{e}rot resonances or interference effects similar to those observed in conventional ultrathin dielectric-metal structures \cite{zhao2021flexible, kats2013nanometre}. $VO_{2}$/Al/S behaves as an asymmetric Fabry-P\'{e}rot cavity leading to the two major hotspots around 500 to 600 nm wavelength range. In order to examine the potential for a color change, the reflectance spectral shift for different thicknesses of $VO_{2}$ at RT and HT is analysed. Fig. \ref{simulation} (e) and \ref{simulation} (f) show the calculated reflectance spectra for 30 nm, 40 nm, and 50 nm of $VO_{2}$ thickness for the structures illustrated in \ref{simulation} (a) and \ref{simulation} (b), respectively. Compared to $VO_{2}$/S, $VO_{2}$/Al/S shows a significant shift in the spectral position of the reflectance minima with changes in temperature and thicknesses of $VO_{2}$. A substantial change in color or chromaticity could be observed for $VO_{2}$/Al/S due to these spectral shifts. As the observed strength of reflectance change is similar for the two hotspot points, we have chosen 40 nm of $VO_{2}$ for our experiments due to the lower thickness advantage.\\

\section{Experimental Details}\label{Experimental Details}
\begin{figure}[H]
    \centering
    \includegraphics[scale=0.55]{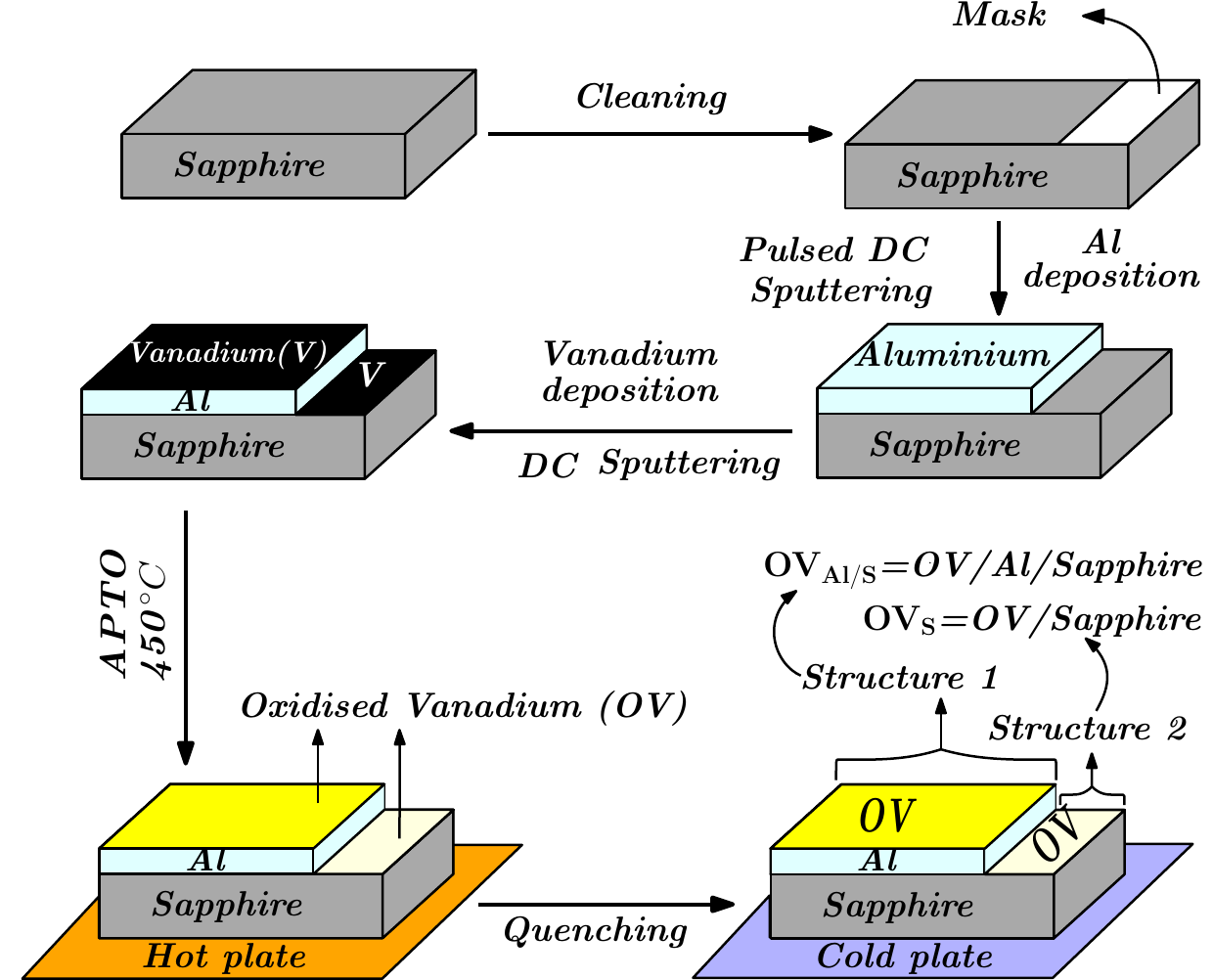}
    \caption{Fabrication process for Oxidised Vanadium (OV)/Al/Sapphire \& OV/Sapphire structures.}
    \label{fig:fabprocessVO2ALpdc350C}
\end{figure}

The process flow followed to fabricate the samples is shown in Fig. \ref{fig:fabprocessVO2ALpdc350C}. First, the substrate (c-plane Sapphire) is cleaned with acetone and isopropyl alcohol in order to prepare for the depositions. For Al deposition, pulsed DC magnetron sputtering is utilized. Prior to deposition, a small mask was placed on the substrate to prevent the deposition of Al in that area. The sputtering power is kept at 30 W. Al films are deposited from a 2-inch diameter Al target for $\sim$ 30 minutes to achieve Al thickness of $\sim$ 50 nm. After unloading the sample, the mask is removed. Following that, thin metallic V films are deposited on Al and sapphire at room temperature by DC magnetron sputtering. A 4-inch diameter V target of 99.95\% purity is used. The DC power is maintained at 90 W. The timing of the deposition is adjusted to achieve a V thickness of $\sim$ 20 nm (which will result in a $VO_{2}$ film of 40 nm thickness after oxidation \cite{gurvitch2007vo}).\\ To oxidize the samples, atmospheric pressure thermal oxidation (APTO) \cite{ashok2020vanadium} is used  which involves oxidation of V in open atmosphere. The samples are placed one by one on a hot plate maintained at 450$^{\circ}$C for different oxidation times. After the desired oxidation duration ($t_{oxd}$), they are quickly cooled to room temperature using a cold plate. Nine samples (labeled $S_{1}-S_{9}$) are prepared with varying $t_{oxd}$ ranging from 6 seconds to 5 minutes, and as a result, the structure of $OV_{Al/S}$ (oxidised vanadium (OV)/Al/sapphire) and $OV_{S}$ (OV/sapphire) is achieved on the same sample for each oxidation duration. For structural characterization of the samples, Raman spectroscopy is performed using an excitation laser of wavelength 532 nm (Acton research corporation spectra pro 2500i). Visible optical reflectance is measured using Filmmetrics F20 spectrometer.

\section{Results and Discussion}\label{Results and Discussion}
\begin{figure}[H]
\centerline{\includegraphics[scale=0.55]{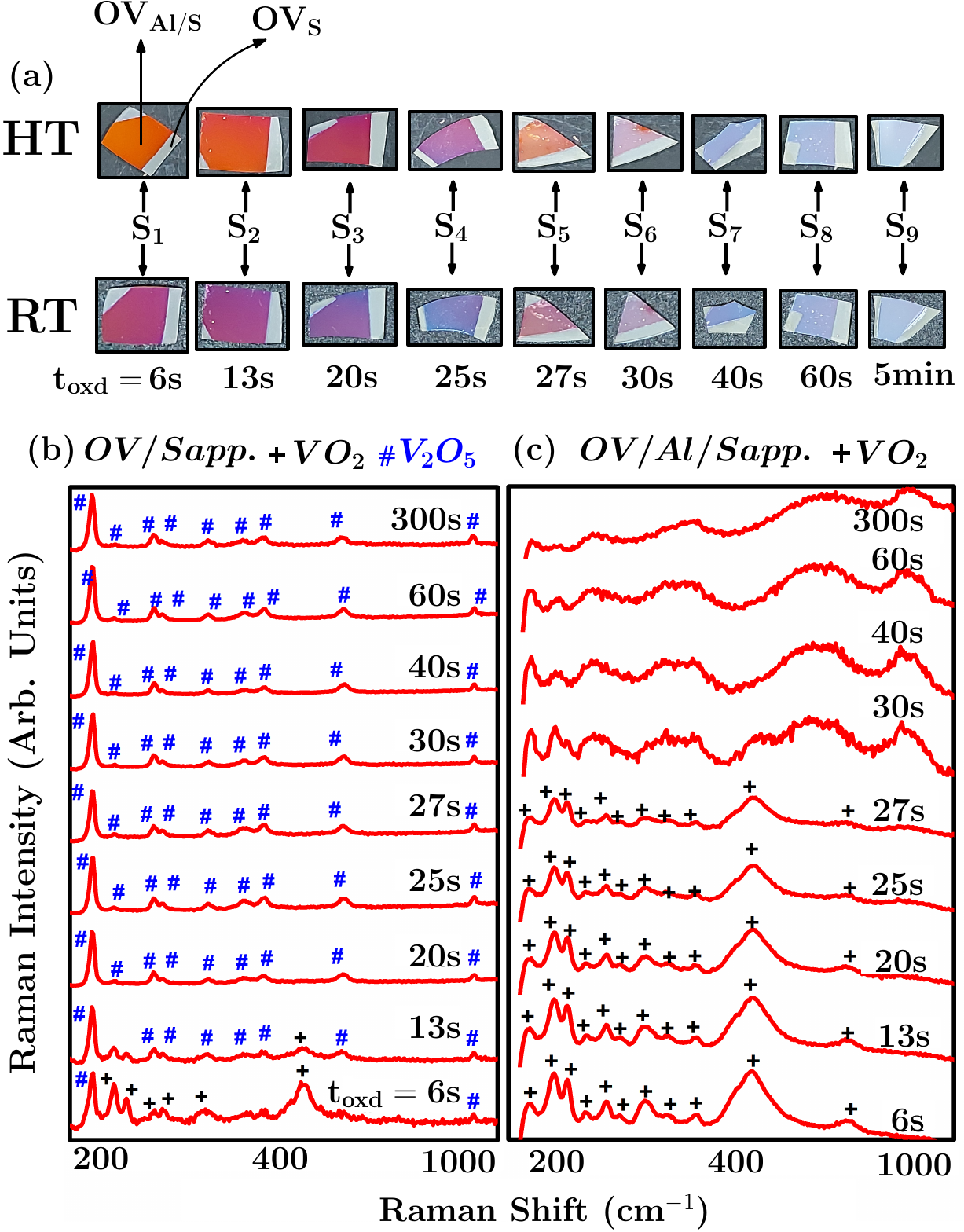}}
\caption{(a) Optical images of all the fabricated samples at RT and HT. Raman characterization results for (b) $OV_{S}$ and (c) $OV_{Al/S}$ structure. The Raman peak positions for $VO_{2}$ and $V_{2}O_{5}$ are referred from \cite{shvets2019review}.}
\label{RamanVO2onAlSapp450C_PDC}
\end{figure}

\begin{figure}[H]
\centerline{\includegraphics[scale=0.55]{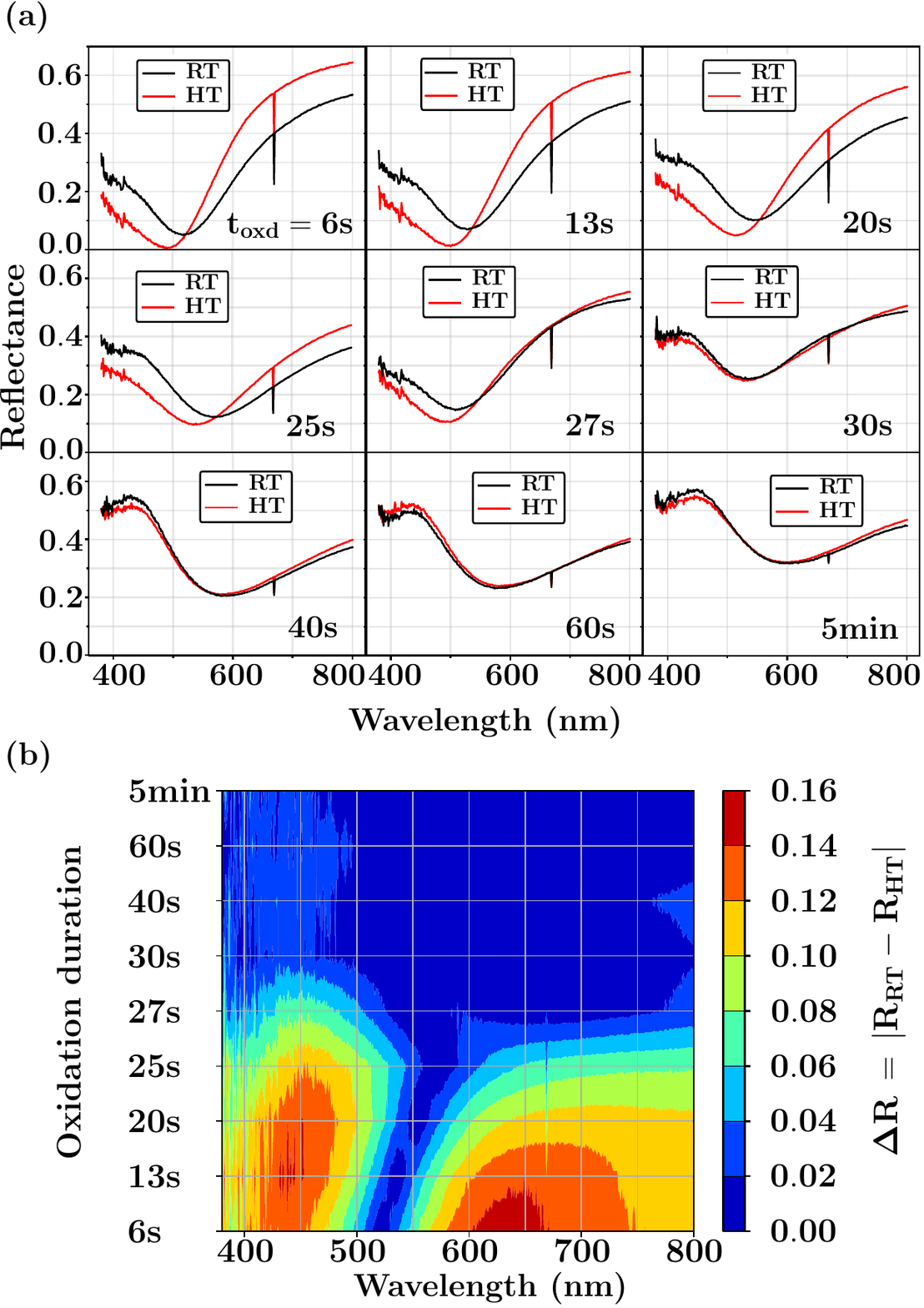}}
\caption{(a) Measured reflectance spectra of $OV_{Al/S}$ samples measured at room and high temperatures for increasing oxidation duration. (b) Reflectance switching ($\Delta R$) comparison curve for $OV_{Al/S}$; $\Delta R$ is calculated by taking absolute difference of measured reflectance at RT and HT.}
\label{RefVO2onAlSapp450C_PDC}
\end{figure}

\begin{figure}[H]
\centerline{\includegraphics[scale = 0.85]{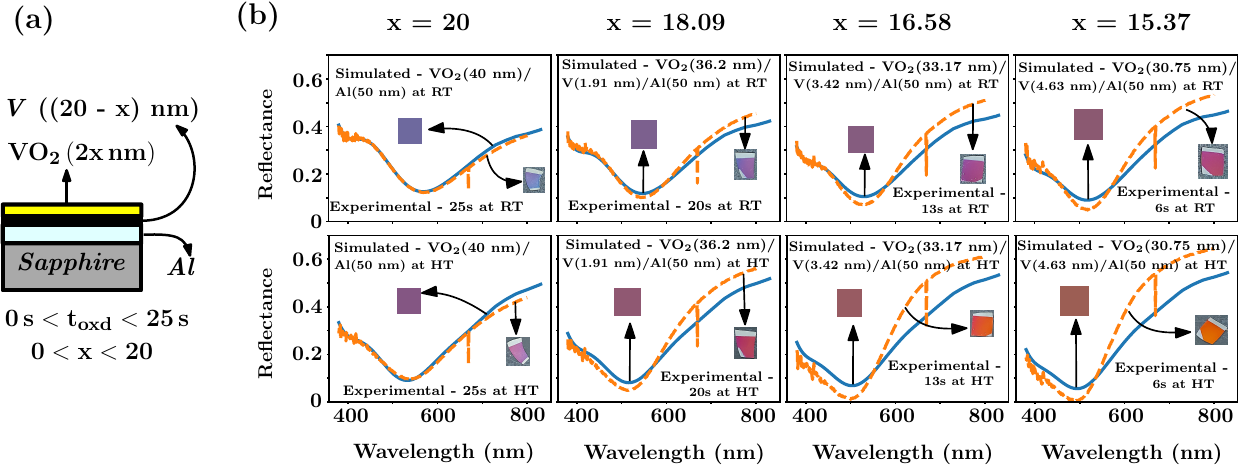}}
\caption{The temporal evolution of oxidation gives rise to relative thickness variations within the structure. The different oxidation duration ($t_{oxd}$) cases modelled as (a) $VO_{2}$ (2x nm)/V ((20 - x) nm)/Al (50 nm)/Sapphire structure for x $\in$ [0,20], due to partial oxidation of V \cite{ashok2022multi}. The zero-thickness point (x = 0) of $VO_{2}$ represents the unoxidized sample V (20 nm)/Al (50 nm)/Sapphire, and the 40 nm point (x = 20) signifies the completely oxidized Vanadium, representing $VO_{2}$ (40 nm)/Al (50 nm)/Sapphire structure. (b) Comparative plots of simulated reflectance and color rendering results alongside experimental data. A reasonable match is observed between the two datasets for the predicted partially oxidized structure.}
\label{PatialOXDVO2onVAlSapp450C_PDC}
\end{figure}

\begin{figure}[H]
\centerline{\includegraphics[scale = 0.5]{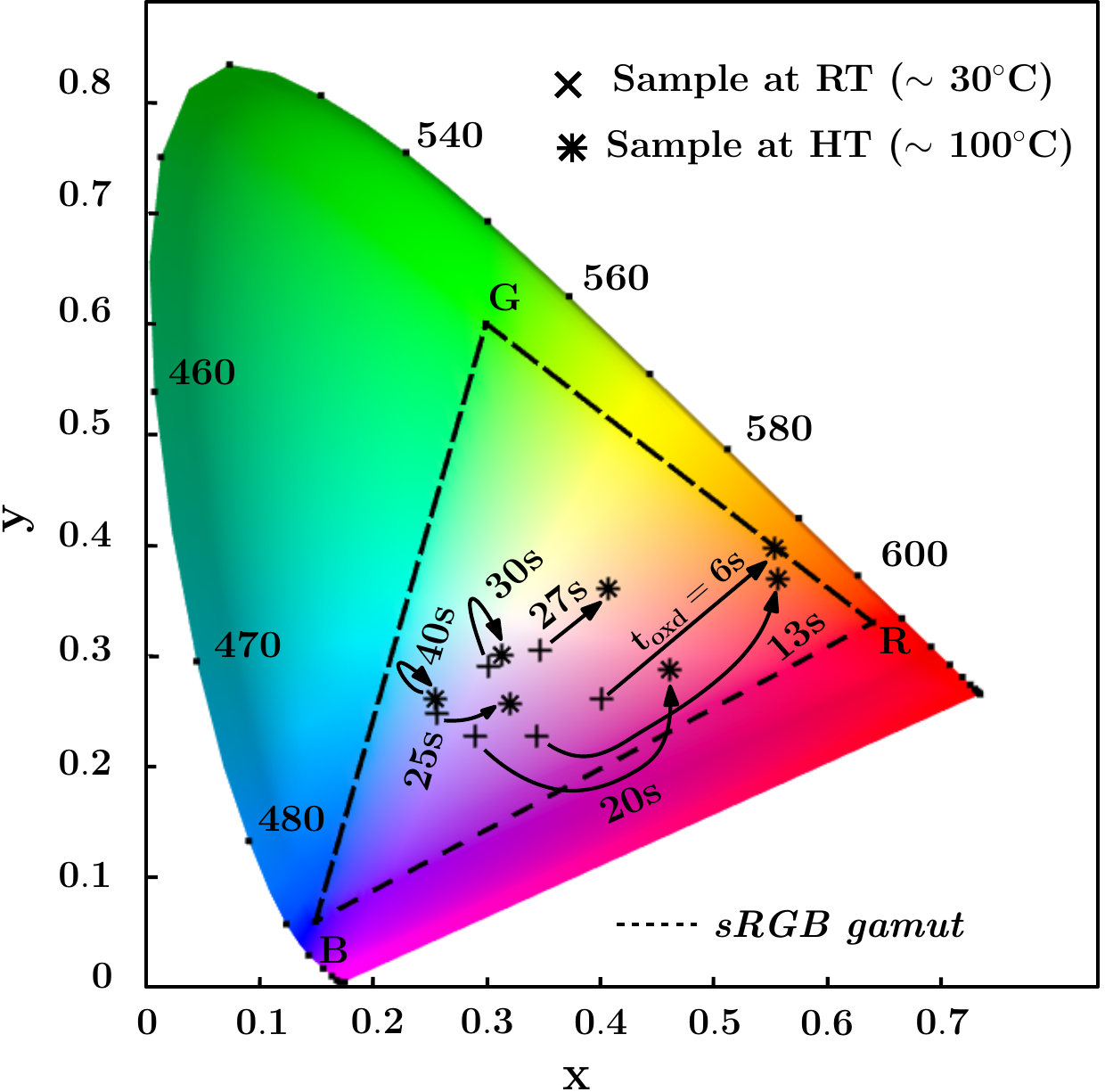}}
\caption{Color mapping of the $OV_{Al/S}$ samples on the CIE 1931-xy chromaticity diagram. Connecting arrow directs from the RT to HT chromaticity points. The color change can be observed for samples $S_{1}$-$S_{5}$ ($t_{oxd}$: 6-27 s); $S_{6}$ ($t_{oxd}$: 30 s) shows very less change in chromaticity. There is no discernible color change in the $S_{7}$ ($t_{oxd}$: 40 s), $S_{8}$ ($t_{oxd}$: 60 s), and $S_{9}$ ($t_{oxd}$: 5 min.) samples; hence, the latter two are not indicated in the plot.}
\label{ColorgamutVO2onAlSapp450C_PDC}
\end{figure}

After fabrication, the effect of temperature on the color of the fabricated samples is analyzed. Camera images are taken while each sample is kept on the heating stage, initially maintained at RT ($\sim$ 30$^{\circ}$C) and subsequently at HT ($\sim$ 100$^{\circ}$C). Figure \ref{RamanVO2onAlSapp450C_PDC} (a) shows the RT and HT optical images for all the fabricated samples. The colored part of the samples corresponds to the $OV_{Al/S}$ structure, and the rectangular, silver-whitish part is $OV_{S}$. 
The samples are cycled multiple times between RT and HT, and the observed colors are found to be consistent and reversible. From Fig. \ref{RamanVO2onAlSapp450C_PDC} (a), it can be seen that the $OV_{S}$ area does not show any color change with temperature. However, for the $OV_{Al/S}$, the 6 s to 27 s oxidation duration ($t_{oxd}$)  samples ($S_{1}-S_{5}$) show significantly different colors at the two temperatures. The samples with $t_{oxd}$ of 30 s to 5 minutes ($S_{6}-S_{9}$) show no color change with temperature. The reversible color-switching videos of the 6 s and 27 s samples (between RT and HT) are included as supplementary material.\\
To understand the observed color-changing behavior, the samples are characterized using Raman and reflectance measurements. Fig. \ref{RamanVO2onAlSapp450C_PDC} (b) and \ref{RamanVO2onAlSapp450C_PDC} (c) show the Raman characterization results for the $OV_{S}$ and $OV_{Al/S}$ structures, respectively. For $OV_{S}$ (Fig. \ref{RamanVO2onAlSapp450C_PDC} (b)), the $VO_{2}$ Raman peaks are observed for $S_{1}$ and $S_{2}$ ($t_{oxd}$: 6 s and 13 s), and for higher $t_{oxd}$ $V_{2}O_{5}$ Raman peaks are detected. This behaviour is consistent with existing studies where the oxidized samples of V/Sapphire report $VO_{2}$ formation followed by $V_{2}O_{5}$ formation \cite{ashok2020vanadium, rampelberg2015situ}. For $OV_{Al/S}$ (Fig. \ref{RamanVO2onAlSapp450C_PDC} (c)), $VO_{2}$ is present in samples $S_{1}$ to $S_{5}$ ($t_{oxd}$: 6-27 s); from $S_{6}$ ($t_{oxd}$: 30 s) onwards, no precise peaks of vanadium oxides are observed till 5 minutes of oxidation. The presence of the underneath Al layer might account for the variation in oxidation behaviour between the two structures. Comparing Fig. \ref{RamanVO2onAlSapp450C_PDC} (b) and \ref{RamanVO2onAlSapp450C_PDC} (c), it is evident that the rate of oxidation is slower in the $OV_{Al/S}$ samples than in $OV_{S}$. The wide peaks for samples $S_{6}$-$S_{9}$ ($t_{oxd}$: 30 s - 5 minutes) in $OV_{Al/S}$  could be attributed to the presence of amorphous vanadium oxides ($VO_{x}$) \cite{shvets2019review, allabergenov2021control, shvets2021correlation}. By correlating the observed Raman spectra with the corresponding sample's optical images, it can be concluded that, for $OV_{Al/S}$, only the $VO_{2}$ -containing samples show color switching; others in which $VO_{x}$ is present do not show visible color changes. In the case of $OV_{S}$, the inherently low optical switching for $VO_{2}$/sapphire and the presence of $V_{2}O_{5}$ could be the reasons for no perceivable difference in visible color with temperature.\\
Next, the reflectance measurement is performed on the $OV_{Al/S}$ samples for the two temperatures RT and HT. Fig. \ref{RefVO2onAlSapp450C_PDC} (a) shows the measured spectra in the visible region for all the samples. Samples $S_{1}$-$S_{5}$ ($t_{oxd}$: 6-27 s) that contain $VO_{2}$ show a significant change in reflectance for the two temperatures. However, negligible change in reflectance is observed for $VO_{x}$ -containing samples $S_{6}$-$S_{9}$ ($t_{oxd}$: 30 s to 5 min.). As discussed, the broad peaks and dip in the reflectance spectra arise due to the Fabry-P\'{e}rot resonances or interference effects in ultrathin film for dielectric-metal structures \cite{zhao2021flexible, kats2013nanometre}. From Fig. \ref{RefVO2onAlSapp450C_PDC} (a), it can be inferred that for samples $S_{1}$-$S_{5}$ ($t_{oxd}$: 6-27 s), the change in color with varying $t_{oxd}$ and temperature is brought about by the spectral shift in reflectance minima. This gradual shift in reflectance with varying $t_{oxd}$ might be accredited to the partial oxidation of V \cite{ashok2022multi}, while the spectral shift with temperature is caused by the phase transition in $VO_{2}$. The sharp drop around 669 nm in the measured reflectance spectra for all the samples (Fig. \ref{RefVO2onAlSapp450C_PDC} (a)) could be due to characteristic spectrum line of the Deuterium lamp present in the Filmmetrics F20 spectrometer. The $\Delta R$ for different wavelengths with varying $t_{oxd}$ is calculated from measured reflectance of Fig. \ref{RefVO2onAlSapp450C_PDC} (a) and plotted in Fig. \ref{RefVO2onAlSapp450C_PDC} (b) using the “contourf” function in Python. This function uses linear interpolation to get points around label. For $VO_{2}$-containing samples $S_{1}$-$S_{5}$ ($t_{oxd}$ = 6 s to 27 s), the trend shows a decrease in $\Delta R$ as $t_{oxd}$ increases. Samples $S_{6}$-$S_{9}$ ($t_{oxd}$ = 30 s to 5 min.) exhibit a minimal change in $\Delta R$, primarily due to the presence of amorphous vanadium oxide. As can be seen from Fig. \ref{RefVO2onAlSapp450C_PDC} (b), a region of minima for $\Delta R$ exists in the wavelength range of 500 to 600 nm, which occurs as the measured reflectance curves at RT and HT cross each other. Compared to this, the calculated $\Delta R$ in Fig. \ref{simulation} (d) shows a region of maxima for the abovementioned wavelength range. This contrasting behavior could be attributed to the substantial difference in the complex refractive index of $VO_{2}$ resulting from different synthesis methods \cite{wan2019optical, barimah2022infrared}.\\
To explain the observed color variation among the $S_{1}$-$S_{4}$ samples exhibiting similar Raman spectra, the different $t_{oxd}$ cases are modeled as $VO_{2}$ (2x nm)/V ((20 - x) nm)/Al (50 nm)/Sapphire structure for x $\in$ [0,20], due to partial oxidation of V \cite{ashok2022multi}, as shown in Fig. \ref{PatialOXDVO2onVAlSapp450C_PDC} (a). The modeled reflectance and the calculated colors are compared against the experimental data. The refractive indices of V and $VO_{2}$ (at RT and HT ) are extracted from the unoxidized sample (20 nm V/ 50 nm Al/ Sapphire) and $t_{oxd}$ = 25 s sample (40 nm $VO_{2}$/50 nm Al/ Sapphire) respective reflectance spectra. The details of the extraction of the refractive indices are included in the supplementary material. Results for the reflectance and color match are shown in Fig. \ref{PatialOXDVO2onVAlSapp450C_PDC} (b). A good fit with all the measured spectra/observed colors is obtained, supporting the partial oxidation assumption. The slight mismatch in the modelled/measured reflectance spectra could be due to the evolution in refractive index with $t_{oxd}$, as the $VO_{2}$ refractive indices depend on various other parameters, such as surface roughness, vertical gradients in microstructure, non-stoichiometry, and film thickness \cite{wan2019optical, motyka2012microstructural, cueff2020vo2}. Extraction of $VO_{2}$ indices from the $S_{5}$ sample (assuming it to be completely oxidized (x = 20)) does not give a good fit to the RT and HT reflectance spectra for all the color-changing samples, possibly due to nucleation of the amorphous $VO_{x}$ phase in this sample.\\
The samples are mapped on the CIE 1931-xy chromaticity diagram, as shown in Fig. \ref{ColorgamutVO2onAlSapp450C_PDC}  (details of the procedure for mapping the reflectance spectra to the chromaticity diagram are included in the supplementary material). This graphical representation indicates the chromaticity coverage of the samples, also called the color gamut. Standard RGB space (sRGB) \cite{susstrunk1999standard} is highlighted along with samples for relative comparison. The arrows shown in the figure are directed from RT to HT chromaticity points. As can be observed, samples $S_{1}$-$S_{5}$ show a significant change in chromaticity due to the $VO_{2}$ phase transition; for $S_{6}$, very little change in chromaticity is observed. No perceptible change in colour is observed for samples $S_{7}$-$S_{9}$; hence among these, only $S_{7}$ is highlighted on the graph. The RT and HT chromaticity points for $S_{7}$ show complete overlap. The experimental samples cover a large color gamut. Many more colors may be observed with suitable tuning in process parameters like the deposition process and oxidation temperature as they directly affect the $VO_{2}$ phase transition.\\ 
As illustrated in Table \ref{conclusion_comparision_table}, this study holds an advantage over previous studies owing to its simple two-layer uniform structure, obviating the need for nano-scale patterning. Compared to Au or Ag, the use of Al reduces the cost. In addition, the thermal budget of the fabrication process is relatively low compared to alternative approaches. However, a limitation of this work lies in the need for precise control of the oxidation time. Controlling such brief $t_{oxd}$ can be challenging; one potential alternative is to lower the oxidation temperature to achieve longer $t_{oxd}$ without significantly compromising the film quality \cite{ashok2023low}. Even though the amorphous vanadium oxide on Al does not show significant color change, its presence and the slower V oxidation rate on Al are worth exploring further through a comprehensive material characterization.\\
\section{Conclusion}\label{Conclusion}
This work demonstrates strong dynamic color switching with $VO_{2}$/Al/sapphire ultrathin ($\sim$ $\frac{\lambda}{14}$) asymmetric Fabry-P\'{e}rot-type thermochromic structure. Consistent and reversible color changes are observed with the change in temperature from low (30$^\circ$C) to high (100$^\circ$C) or vice versa. Multiple colours are observed for various oxidation durations. The oxidized samples show a significant change in chromaticity with the phase change of $VO_{2}$, thereby covering a wide color gamut. Compared to other works in a similar area, the structure considered in this work is simple, uniform, and does not require complex patterning. This work can significantly contribute to the design of simple and reversible dynamic/active color structures. This work covers the red and blue-violet regions, however green color region is not covered well. Effectively reproducing all three primary colors could potentially lead to its application in the fields of display and visible camouflage.\\
\section*{Supplementary Material}
See the supplementary material for the details of Aluminium thickness optimisation, extraction of refractive indices of V, $VO_{2}$ (RT) and $VO_{2}$ (HT), calculation of x and y color gamut parameters from the reflectance spectra. Additionally, the reversible color-switching videos of the $S_{1}$ and $S_{5}$ samples along with their RGB variation during the color transition of sample (between RT and HT) is also included.\\
\section{Acknowledgments}\label{acknowledgments}
This project was supported by Science and Engineering Research Board, India via Imprint Grant No. IMP/2018/000404 and Core Research Grant No. CRG/2022/005421.\\
\section*{Conflict of Interest}\label{Conflict of Interest}
The authors declare no competing financial interest. 

%
\bibliography{achemso-demo}

\providecommand{\latin}[1]{#1}
\makeatletter
\providecommand{\doi}
  {\begingroup\let\do\@makeother\dospecials
  \catcode`\{=1 \catcode`\}=2 \doi@aux}
\providecommand{\doi@aux}[1]{\endgroup\texttt{#1}}
\makeatother
\providecommand*\mcitethebibliography{\thebibliography}
\csname @ifundefined\endcsname{endmcitethebibliography}
  {\let\endmcitethebibliography\endthebibliography}{}
\begin{mcitethebibliography}{41}
\providecommand*\natexlab[1]{#1}
\providecommand*\mciteSetBstSublistMode[1]{}
\providecommand*\mciteSetBstMaxWidthForm[2]{}
\providecommand*\mciteBstWouldAddEndPuncttrue
  {\def\EndOfBibitem{\unskip.}}
\providecommand*\mciteBstWouldAddEndPunctfalse
  {\let\EndOfBibitem\relax}
\providecommand*\mciteSetBstMidEndSepPunct[3]{}
\providecommand*\mciteSetBstSublistLabelBeginEnd[3]{}
\providecommand*\EndOfBibitem{}
\mciteSetBstSublistMode{f}
\mciteSetBstMaxWidthForm{subitem}{(\alph{mcitesubitemcount})}
\mciteSetBstSublistLabelBeginEnd
  {\mcitemaxwidthsubitemform\space}
  {\relax}
  {\relax}

\bibitem[Wang \latin{et~al.}(2016)Wang, Runnerstrom, and
  Milliron]{doi:10.1146/annurev-chembioeng-080615-034647}
Wang,~Y.; Runnerstrom,~E.~L.; Milliron,~D.~J. Switchable Materials for Smart
  Windows. \emph{Annual Review of Chemical and Biomolecular Engineering}
  \textbf{2016}, \emph{7}, 283--304, PMID: 27023660\relax
\mciteBstWouldAddEndPuncttrue
\mciteSetBstMidEndSepPunct{\mcitedefaultmidpunct}
{\mcitedefaultendpunct}{\mcitedefaultseppunct}\relax
\EndOfBibitem
\bibitem[Ke \latin{et~al.}(2016)Ke, Balin, Wang, Lu, Tok, White, Magdassi,
  Abdulhalim, and Long]{ke2016two}
Ke,~Y.; Balin,~I.; Wang,~N.; Lu,~Q.; Tok,~A. I.~Y.; White,~T.~J.; Magdassi,~S.;
  Abdulhalim,~I.; Long,~Y. Two-dimensional SiO2/VO2 photonic crystals with
  statically visible and dynamically infrared modulated for smart window
  deployment. \emph{ACS applied materials \& interfaces} \textbf{2016},
  \emph{8}, 33112--33120\relax
\mciteBstWouldAddEndPuncttrue
\mciteSetBstMidEndSepPunct{\mcitedefaultmidpunct}
{\mcitedefaultendpunct}{\mcitedefaultseppunct}\relax
\EndOfBibitem
\bibitem[Kepic \latin{et~al.}(2021)Kepic, Ligmajer, Hrton, Ren, Menezes, Maier,
  and Sikola]{kepic2021optically}
Kepic,~P.; Ligmajer,~F.; Hrton,~M.; Ren,~H.; Menezes,~L. d.~S.; Maier,~S.~A.;
  Sikola,~T. Optically tunable Mie resonance VO2 nanoantennas for metasurfaces
  in the visible. \emph{ACS Photonics} \textbf{2021}, \emph{8},
  1048--1057\relax
\mciteBstWouldAddEndPuncttrue
\mciteSetBstMidEndSepPunct{\mcitedefaultmidpunct}
{\mcitedefaultendpunct}{\mcitedefaultseppunct}\relax
\EndOfBibitem
\bibitem[He \latin{et~al.}(2020)He, Youngblood, Cheng, Miao, and
  Bhaskaran]{he2020dynamically}
He,~Q.; Youngblood,~N.; Cheng,~Z.; Miao,~X.; Bhaskaran,~H. Dynamically tunable
  transmissive color filters using ultra-thin phase change materials.
  \emph{Optics Express} \textbf{2020}, \emph{28}, 39841--39849\relax
\mciteBstWouldAddEndPuncttrue
\mciteSetBstMidEndSepPunct{\mcitedefaultmidpunct}
{\mcitedefaultendpunct}{\mcitedefaultseppunct}\relax
\EndOfBibitem
\bibitem[Duan \latin{et~al.}(2017)Duan, Kamin, and Liu]{duan2017dynamic}
Duan,~X.; Kamin,~S.; Liu,~N. Dynamic plasmonic colour display. \emph{Nature
  communications} \textbf{2017}, \emph{8}, 14606\relax
\mciteBstWouldAddEndPuncttrue
\mciteSetBstMidEndSepPunct{\mcitedefaultmidpunct}
{\mcitedefaultendpunct}{\mcitedefaultseppunct}\relax
\EndOfBibitem
\bibitem[Garshasbi and Santamouris(2019)Garshasbi, and
  Santamouris]{garshasbi2019using}
Garshasbi,~S.; Santamouris,~M. Using advanced thermochromic technologies in the
  built environment: Recent development and potential to decrease the energy
  consumption and fight urban overheating. \emph{Solar Energy Materials and
  Solar Cells} \textbf{2019}, \emph{191}, 21--32\relax
\mciteBstWouldAddEndPuncttrue
\mciteSetBstMidEndSepPunct{\mcitedefaultmidpunct}
{\mcitedefaultendpunct}{\mcitedefaultseppunct}\relax
\EndOfBibitem
\bibitem[Chain(1991)]{chain1991optical}
Chain,~E.~E. Optical properties of vanadium dioxide and vanadium pentoxide thin
  films. \emph{Applied optics} \textbf{1991}, \emph{30}, 2782--2787\relax
\mciteBstWouldAddEndPuncttrue
\mciteSetBstMidEndSepPunct{\mcitedefaultmidpunct}
{\mcitedefaultendpunct}{\mcitedefaultseppunct}\relax
\EndOfBibitem
\bibitem[Wu \latin{et~al.}(2011)Wu, Zimmers, Aubin, Ghosh, Liu, and
  Lopez]{wu2011electric}
Wu,~B.; Zimmers,~A.; Aubin,~H.; Ghosh,~R.; Liu,~Y.; Lopez,~R.
  Electric-field-driven phase transition in vanadium dioxide. \emph{Physical
  Review B} \textbf{2011}, \emph{84}, 241410\relax
\mciteBstWouldAddEndPuncttrue
\mciteSetBstMidEndSepPunct{\mcitedefaultmidpunct}
{\mcitedefaultendpunct}{\mcitedefaultseppunct}\relax
\EndOfBibitem
\bibitem[Kawakubo and Nakagawa(1964)Kawakubo, and Nakagawa]{kawakubo1964phase}
Kawakubo,~T.; Nakagawa,~T. Phase transition in VO2. \emph{Journal of the
  Physical Society of Japan} \textbf{1964}, \emph{19}, 517--519\relax
\mciteBstWouldAddEndPuncttrue
\mciteSetBstMidEndSepPunct{\mcitedefaultmidpunct}
{\mcitedefaultendpunct}{\mcitedefaultseppunct}\relax
\EndOfBibitem
\bibitem[Liu \latin{et~al.}(2012)Liu, Hwang, Tao, Strikwerda, Fan, Keiser,
  Sternbach, West, Kittiwatanakul, Lu, \latin{et~al.} others]{liu2012terahertz}
Liu,~M.; Hwang,~H.~Y.; Tao,~H.; Strikwerda,~A.~C.; Fan,~K.; Keiser,~G.~R.;
  Sternbach,~A.~J.; West,~K.~G.; Kittiwatanakul,~S.; Lu,~J.; others
  Terahertz-field-induced insulator-to-metal transition in vanadium dioxide
  metamaterial. \emph{Nature} \textbf{2012}, \emph{487}, 345--348\relax
\mciteBstWouldAddEndPuncttrue
\mciteSetBstMidEndSepPunct{\mcitedefaultmidpunct}
{\mcitedefaultendpunct}{\mcitedefaultseppunct}\relax
\EndOfBibitem
\bibitem[Cao \latin{et~al.}(2009)Cao, Ertekin, Srinivasan, Fan, Huang, Zheng,
  Yim, Khanal, Ogletree, Grossman, \latin{et~al.} others]{cao2009strain}
Cao,~J.; Ertekin,~E.; Srinivasan,~V.; Fan,~W.; Huang,~S.; Zheng,~H.; Yim,~J.;
  Khanal,~D.; Ogletree,~D.; Grossman,~J.; others Strain engineering and
  one-dimensional organization of metal--insulator domains in single-crystal
  vanadium dioxide beams. \emph{Nature nanotechnology} \textbf{2009}, \emph{4},
  732--737\relax
\mciteBstWouldAddEndPuncttrue
\mciteSetBstMidEndSepPunct{\mcitedefaultmidpunct}
{\mcitedefaultendpunct}{\mcitedefaultseppunct}\relax
\EndOfBibitem
\bibitem[De~Natale \latin{et~al.}(1989)De~Natale, Hood, and
  Harker]{de1989formation}
De~Natale,~J.~F.; Hood,~P.; Harker,~A.~B. Formation and characterization of
  grain-oriented VO2 thin films. \emph{Journal of Applied Physics}
  \textbf{1989}, \emph{66}, 5844--5850\relax
\mciteBstWouldAddEndPuncttrue
\mciteSetBstMidEndSepPunct{\mcitedefaultmidpunct}
{\mcitedefaultendpunct}{\mcitedefaultseppunct}\relax
\EndOfBibitem
\bibitem[Jin \latin{et~al.}(1997)Jin, Yoshimura, and
  Tanemura]{jin1997dependence}
Jin,~P.; Yoshimura,~K.; Tanemura,~S. Dependence of microstructure and
  thermochromism on substrate temperature for sputter-deposited VO 2 epitaxial
  films. \emph{Journal of Vacuum Science \& Technology A: Vacuum, Surfaces, and
  Films} \textbf{1997}, \emph{15}, 1113--1117\relax
\mciteBstWouldAddEndPuncttrue
\mciteSetBstMidEndSepPunct{\mcitedefaultmidpunct}
{\mcitedefaultendpunct}{\mcitedefaultseppunct}\relax
\EndOfBibitem
\bibitem[Ashok \latin{et~al.}(2020)Ashok, Chauhan, and Verma]{ashok2020high}
Ashok,~P.; Chauhan,~Y.~S.; Verma,~A. High infrared reflectance modulation in
  VO2 films synthesized on glass and ITO coated glass substrates using
  atmospheric oxidation of vanadium. \emph{Optical Materials} \textbf{2020},
  \emph{110}, 110438\relax
\mciteBstWouldAddEndPuncttrue
\mciteSetBstMidEndSepPunct{\mcitedefaultmidpunct}
{\mcitedefaultendpunct}{\mcitedefaultseppunct}\relax
\EndOfBibitem
\bibitem[Saitzek \latin{et~al.}(2008)Saitzek, Guinneton, Guirleo, Sauques,
  Aguir, and Gavarri]{saitzek2008vo2}
Saitzek,~S.; Guinneton,~F.; Guirleo,~G.; Sauques,~L.; Aguir,~K.; Gavarri,~J.-R.
  VO2 thin films deposited on silicon substrates from V2O5 target: limits in
  optical switching properties and modeling. \emph{Thin Solid Films}
  \textbf{2008}, \emph{516}, 891--897\relax
\mciteBstWouldAddEndPuncttrue
\mciteSetBstMidEndSepPunct{\mcitedefaultmidpunct}
{\mcitedefaultendpunct}{\mcitedefaultseppunct}\relax
\EndOfBibitem
\bibitem[Graf \latin{et~al.}(2017)Graf, Schlafer, Garbe, Klein, and
  Mathur]{graf2017interdependence}
Graf,~D.; Schlafer,~J.; Garbe,~S.; Klein,~A.; Mathur,~S. Interdependence of
  structure, morphology, and phase transitions in CVD grown VO2 and V2O3
  nanostructures. \emph{Chemistry of Materials} \textbf{2017}, \emph{29},
  5877--5885\relax
\mciteBstWouldAddEndPuncttrue
\mciteSetBstMidEndSepPunct{\mcitedefaultmidpunct}
{\mcitedefaultendpunct}{\mcitedefaultseppunct}\relax
\EndOfBibitem
\bibitem[Zhang \latin{et~al.}(2010)Zhang, Gao, Chen, Du, Cao, Kang, and
  Luo]{zhang2010thermochromic}
Zhang,~Z.; Gao,~Y.; Chen,~Z.; Du,~J.; Cao,~C.; Kang,~L.; Luo,~H. Thermochromic
  VO2 thin films: solution-based processing, improved optical properties, and
  lowered phase transformation temperature. \emph{Langmuir} \textbf{2010},
  \emph{26}, 10738--10744\relax
\mciteBstWouldAddEndPuncttrue
\mciteSetBstMidEndSepPunct{\mcitedefaultmidpunct}
{\mcitedefaultendpunct}{\mcitedefaultseppunct}\relax
\EndOfBibitem
\bibitem[Liu \latin{et~al.}(2017)Liu, Su, Kaneti, Chen, Tang, Yuan, Gao, Jiang,
  Jiang, and Yu]{liu2017dual}
Liu,~M.; Su,~B.; Kaneti,~Y.~V.; Chen,~Z.; Tang,~Y.; Yuan,~Y.; Gao,~Y.;
  Jiang,~L.; Jiang,~X.; Yu,~A. Dual-phase transformation: Spontaneous
  self-template surface-patterning strategy for ultra-transparent VO2 solar
  modulating coatings. \emph{ACS nano} \textbf{2017}, \emph{11}, 407--415\relax
\mciteBstWouldAddEndPuncttrue
\mciteSetBstMidEndSepPunct{\mcitedefaultmidpunct}
{\mcitedefaultendpunct}{\mcitedefaultseppunct}\relax
\EndOfBibitem
\bibitem[Khaled and Berardi(2021)Khaled, and Berardi]{khaled2021current}
Khaled,~K.; Berardi,~U. Current and future coating technologies for
  architectural glazing applications. \emph{Energy and Buildings}
  \textbf{2021}, \emph{244}, 111022\relax
\mciteBstWouldAddEndPuncttrue
\mciteSetBstMidEndSepPunct{\mcitedefaultmidpunct}
{\mcitedefaultendpunct}{\mcitedefaultseppunct}\relax
\EndOfBibitem
\bibitem[Song \latin{et~al.}(2018)Song, Ma, Pu, Li, Guo, Gao, and
  Luo]{song2018tailoring}
Song,~S.; Ma,~X.; Pu,~M.; Li,~X.; Guo,~Y.; Gao,~P.; Luo,~X. Tailoring active
  color rendering and multiband photodetection in a vanadium-dioxide-based
  metamaterial absorber. \emph{Photonics Research} \textbf{2018}, \emph{6},
  492--497\relax
\mciteBstWouldAddEndPuncttrue
\mciteSetBstMidEndSepPunct{\mcitedefaultmidpunct}
{\mcitedefaultendpunct}{\mcitedefaultseppunct}\relax
\EndOfBibitem
\bibitem[Wilson \latin{et~al.}(2018)Wilson, Marocico, and
  Bradley]{wilson2018dynamic}
Wilson,~K.; Marocico,~C.; Bradley,~A. Dynamic structural colour using vanadium
  dioxide thin films. \emph{Journal of Physics D: Applied Physics}
  \textbf{2018}, \emph{51}, 255101\relax
\mciteBstWouldAddEndPuncttrue
\mciteSetBstMidEndSepPunct{\mcitedefaultmidpunct}
{\mcitedefaultendpunct}{\mcitedefaultseppunct}\relax
\EndOfBibitem
\bibitem[Shu \latin{et~al.}(2018)Shu, Yu, Peng, Zhu, Xiong, Fan, Wang, Liu, and
  Wang]{shu2018dynamic}
Shu,~F.-Z.; Yu,~F.-F.; Peng,~R.-W.; Zhu,~Y.-Y.; Xiong,~B.; Fan,~R.-H.;
  Wang,~Z.-H.; Liu,~Y.; Wang,~M. Dynamic plasmonic color generation based on
  phase transition of vanadium dioxide. \emph{Advanced Optical Materials}
  \textbf{2018}, \emph{6}, 1700939\relax
\mciteBstWouldAddEndPuncttrue
\mciteSetBstMidEndSepPunct{\mcitedefaultmidpunct}
{\mcitedefaultendpunct}{\mcitedefaultseppunct}\relax
\EndOfBibitem
\bibitem[Zhou \latin{et~al.}(2018)Zhou, Hu, Song, Li, Qiang, and
  Liang]{zhou2018modified}
Zhou,~L.; Hu,~M.; Song,~X.; Li,~P.; Qiang,~X.; Liang,~J. Modified color for
  VO2/Au/VO2 sandwich structure-based smart windows. \emph{Applied Physics A}
  \textbf{2018}, \emph{124}, 1--6\relax
\mciteBstWouldAddEndPuncttrue
\mciteSetBstMidEndSepPunct{\mcitedefaultmidpunct}
{\mcitedefaultendpunct}{\mcitedefaultseppunct}\relax
\EndOfBibitem
\bibitem[In \latin{et~al.}(2020)In, Cho, Park, Kim, Kim, Noh, and
  Park]{in2020self}
In,~S.; Cho,~J.; Park,~J.; Kim,~S.~Y.; Kim,~H.-T.; Noh,~T.~W.; Park,~N.
  Self-Organized Gold Network--Vanadium Dioxide Hybrid Film for Dynamic
  Modulation of Visible-to-Near-Infrared Light. \emph{Advanced Photonics
  Research} \textbf{2020}, \emph{1}, 2000050\relax
\mciteBstWouldAddEndPuncttrue
\mciteSetBstMidEndSepPunct{\mcitedefaultmidpunct}
{\mcitedefaultendpunct}{\mcitedefaultseppunct}\relax
\EndOfBibitem
\bibitem[Zhao \latin{et~al.}(2021)Zhao, Zhou, Huo, Gao, Ma, and
  Yu]{zhao2021flexible}
Zhao,~J.; Zhou,~Y.; Huo,~Y.; Gao,~B.; Ma,~Y.; Yu,~Y. Flexible dynamic
  structural color based on an ultrathin asymmetric Fabry-Perot cavity with
  phase-change material for temperature perception. \emph{Optics Express}
  \textbf{2021}, \emph{29}, 23273--23281\relax
\mciteBstWouldAddEndPuncttrue
\mciteSetBstMidEndSepPunct{\mcitedefaultmidpunct}
{\mcitedefaultendpunct}{\mcitedefaultseppunct}\relax
\EndOfBibitem
\bibitem[Kats \latin{et~al.}(2013)Kats, Blanchard, Genevet, and
  Capasso]{kats2013nanometre}
Kats,~M.~A.; Blanchard,~R.; Genevet,~P.; Capasso,~F. Nanometre optical coatings
  based on strong interference effects in highly absorbing media. \emph{Nature
  materials} \textbf{2013}, \emph{12}, 20--24\relax
\mciteBstWouldAddEndPuncttrue
\mciteSetBstMidEndSepPunct{\mcitedefaultmidpunct}
{\mcitedefaultendpunct}{\mcitedefaultseppunct}\relax
\EndOfBibitem
\bibitem[Wan \latin{et~al.}(2019)Wan, Zhang, Woolf, Hessel, Rensberg, Hensley,
  Xiao, Shahsafi, Salman, Richter, \latin{et~al.} others]{wan2019optical}
Wan,~C.; Zhang,~Z.; Woolf,~D.; Hessel,~C.~M.; Rensberg,~J.; Hensley,~J.~M.;
  Xiao,~Y.; Shahsafi,~A.; Salman,~J.; Richter,~S.; others On the optical
  properties of thin-film vanadium dioxide from the visible to the far
  infrared. \emph{Annalen der Physik} \textbf{2019}, \emph{531}, 1900188\relax
\mciteBstWouldAddEndPuncttrue
\mciteSetBstMidEndSepPunct{\mcitedefaultmidpunct}
{\mcitedefaultendpunct}{\mcitedefaultseppunct}\relax
\EndOfBibitem
\bibitem[Bbyrnes(2020)]{pythontmm}
Bbyrnes,~S. {Python TMM package}. \url{https://pypi.org/project/tmm/},
  2020\relax
\mciteBstWouldAddEndPuncttrue
\mciteSetBstMidEndSepPunct{\mcitedefaultmidpunct}
{\mcitedefaultendpunct}{\mcitedefaultseppunct}\relax
\EndOfBibitem
\bibitem[Gurvitch \latin{et~al.}(2007)Gurvitch, Luryi, Polyakov, Shabalov,
  Dudley, Wang, Ge, and Yakovlev]{gurvitch2007vo}
Gurvitch,~M.; Luryi,~S.; Polyakov,~A.; Shabalov,~A.; Dudley,~M.; Wang,~G.;
  Ge,~S.; Yakovlev,~V. VO 2 films with strong semiconductor to metal phase
  transition prepared by the precursor oxidation process. \emph{Journal of
  Applied Physics} \textbf{2007}, \emph{102}, 033504\relax
\mciteBstWouldAddEndPuncttrue
\mciteSetBstMidEndSepPunct{\mcitedefaultmidpunct}
{\mcitedefaultendpunct}{\mcitedefaultseppunct}\relax
\EndOfBibitem
\bibitem[Ashok \latin{et~al.}(2020)Ashok, Chauhan, and
  Verma]{ashok2020vanadium}
Ashok,~P.; Chauhan,~Y.~S.; Verma,~A. Vanadium dioxide thin films synthesized
  using low thermal budget atmospheric oxidation. \emph{Thin Solid Films}
  \textbf{2020}, \emph{706}, 138003\relax
\mciteBstWouldAddEndPuncttrue
\mciteSetBstMidEndSepPunct{\mcitedefaultmidpunct}
{\mcitedefaultendpunct}{\mcitedefaultseppunct}\relax
\EndOfBibitem
\bibitem[Shvets \latin{et~al.}(2019)Shvets, Dikaya, Maksimova, and
  Goikhman]{shvets2019review}
Shvets,~P.; Dikaya,~O.; Maksimova,~K.; Goikhman,~A. A review of Raman
  spectroscopy of vanadium oxides. \emph{Journal of Raman spectroscopy}
  \textbf{2019}, \emph{50}, 1226--1244\relax
\mciteBstWouldAddEndPuncttrue
\mciteSetBstMidEndSepPunct{\mcitedefaultmidpunct}
{\mcitedefaultendpunct}{\mcitedefaultseppunct}\relax
\EndOfBibitem
\bibitem[Rampelberg \latin{et~al.}(2015)Rampelberg, De~Schutter, Devulder,
  Martens, Radu, and Detavernier]{rampelberg2015situ}
Rampelberg,~G.; De~Schutter,~B.; Devulder,~W.; Martens,~K.; Radu,~I.;
  Detavernier,~C. In situ X-ray diffraction study of the controlled oxidation
  and reduction in the V--O system for the synthesis of VO 2 and V 2 O 3 thin
  films. \emph{Journal of Materials Chemistry C} \textbf{2015}, \emph{3},
  11357--11365\relax
\mciteBstWouldAddEndPuncttrue
\mciteSetBstMidEndSepPunct{\mcitedefaultmidpunct}
{\mcitedefaultendpunct}{\mcitedefaultseppunct}\relax
\EndOfBibitem
\bibitem[Allabergenov \latin{et~al.}(2021)Allabergenov, Yun, Cho, Lyu, and
  Choi]{allabergenov2021control}
Allabergenov,~B.; Yun,~S.; Cho,~H.-S.; Lyu,~H.-K.; Choi,~B. Control of
  Polymorphic Properties of Multivalent Vanadium Oxide Thin Films. \emph{ACS
  Applied Electronic Materials} \textbf{2021}, \emph{3}, 1142--1150\relax
\mciteBstWouldAddEndPuncttrue
\mciteSetBstMidEndSepPunct{\mcitedefaultmidpunct}
{\mcitedefaultendpunct}{\mcitedefaultseppunct}\relax
\EndOfBibitem
\bibitem[Shvets \latin{et~al.}(2021)Shvets, Maksimova, and
  Goikhman]{shvets2021correlation}
Shvets,~P.; Maksimova,~K.; Goikhman,~A. Correlation between Raman spectra and
  oxygen content in amorphous vanadium oxides. \emph{Physica B: Condensed
  Matter} \textbf{2021}, \emph{613}, 412995\relax
\mciteBstWouldAddEndPuncttrue
\mciteSetBstMidEndSepPunct{\mcitedefaultmidpunct}
{\mcitedefaultendpunct}{\mcitedefaultseppunct}\relax
\EndOfBibitem
\bibitem[Ashok \latin{et~al.}(2022)Ashok, Chauhan, and Verma]{ashok2022multi}
Ashok,~P.; Chauhan,~Y.~S.; Verma,~A. Multi spectral switchable infra-red
  reflectance resonances in highly subwavelength partially oxidized vanadium
  thin films. \emph{Optical Materials} \textbf{2022}, \emph{132}, 112854\relax
\mciteBstWouldAddEndPuncttrue
\mciteSetBstMidEndSepPunct{\mcitedefaultmidpunct}
{\mcitedefaultendpunct}{\mcitedefaultseppunct}\relax
\EndOfBibitem
\bibitem[Barimah \latin{et~al.}(2022)Barimah, Boontan, Steenson, and
  Jose]{barimah2022infrared}
Barimah,~E.~K.; Boontan,~A.; Steenson,~D.~P.; Jose,~G. Infrared optical
  properties modulation of VO2 thin film fabricated by ultrafast pulsed laser
  deposition for thermochromic smart window applications. \emph{Scientific
  Reports} \textbf{2022}, \emph{12}, 11421\relax
\mciteBstWouldAddEndPuncttrue
\mciteSetBstMidEndSepPunct{\mcitedefaultmidpunct}
{\mcitedefaultendpunct}{\mcitedefaultseppunct}\relax
\EndOfBibitem
\bibitem[Motyka \latin{et~al.}(2012)Motyka, Gauntt, Horn, Dickey, and
  Podraza]{motyka2012microstructural}
Motyka,~M.; Gauntt,~B.; Horn,~M.; Dickey,~E.; Podraza,~N. Microstructural
  evolution of thin film vanadium oxide prepared by pulsed-direct current
  magnetron sputtering. \emph{Journal of Applied Physics} \textbf{2012},
  \emph{112}\relax
\mciteBstWouldAddEndPuncttrue
\mciteSetBstMidEndSepPunct{\mcitedefaultmidpunct}
{\mcitedefaultendpunct}{\mcitedefaultseppunct}\relax
\EndOfBibitem
\bibitem[Cueff \latin{et~al.}(2020)Cueff, John, Zhang, Parra, Sun, Orobtchouk,
  Ramanathan, and Sanchis]{cueff2020vo2}
Cueff,~S.; John,~J.; Zhang,~Z.; Parra,~J.; Sun,~J.; Orobtchouk,~R.;
  Ramanathan,~S.; Sanchis,~P. VO2 nanophotonics. \emph{APL Photonics}
  \textbf{2020}, \emph{5}\relax
\mciteBstWouldAddEndPuncttrue
\mciteSetBstMidEndSepPunct{\mcitedefaultmidpunct}
{\mcitedefaultendpunct}{\mcitedefaultseppunct}\relax
\EndOfBibitem
\bibitem[S{\"u}sstrunk \latin{et~al.}(1999)S{\"u}sstrunk, Buckley, and
  Swen]{susstrunk1999standard}
S{\"u}sstrunk,~S.; Buckley,~R.; Swen,~S. Standard RGB color spaces. Color and
  imaging conference. 1999; pp 127--134\relax
\mciteBstWouldAddEndPuncttrue
\mciteSetBstMidEndSepPunct{\mcitedefaultmidpunct}
{\mcitedefaultendpunct}{\mcitedefaultseppunct}\relax
\EndOfBibitem
\bibitem[Ashok \latin{et~al.}(2023)Ashok, Chauhan, and Verma]{ashok2023low}
Ashok,~P.; Chauhan,~Y.~S.; Verma,~A. Low temperature synthesis of VO2 and
  hysteresis free VOx thin films with high temperature coefficient of
  resistance for bolometer applications. \emph{Thin Solid Films} \textbf{2023},
  \emph{781}, 139975\relax
\mciteBstWouldAddEndPuncttrue
\mciteSetBstMidEndSepPunct{\mcitedefaultmidpunct}
{\mcitedefaultendpunct}{\mcitedefaultseppunct}\relax
\EndOfBibitem
\end{mcitethebibliography}

\newpage

\begin{table}[H]
\centering
\caption{Comparison of this work with other $VO_{2}$ based thermochromic studies }
\label{conclusion_comparision_table}
\resizebox{\textwidth}{!}{%
\begin{tabular}{|c|c|c|c|}
\hline
\textbf{Thermochromic Structure}                                                                            & \textbf{Advantages}                                                                                                                 & \textbf{Limitations}   
& \textbf{Ref.}                                                                             \\ \hline
Ag(nano-particle)/insulator/$VO_{2}$/Ag                & \begin{tabular}[c]{@{}c@{}}The presence of nano-particles could lead to more color changes. \end{tabular}                                                                                                                                 &\begin{tabular}[c]{@{}c@{}}Only simulations study. \\ Nanoscale patterning in the structure. \end{tabular}  & \cite{song2018tailoring} \\ \hline  
 ITO/$VO_{2}$/ITO/Ag/Si                                                                                      & \begin{tabular}[c]{@{}c@{}}Different colors are reported by tuning the thicknesses of multiple layers in the structure. \end{tabular} & Only simulations study. & \cite{wilson2018dynamic}                                                                                                        \\ \hline
Ag(nanodisk)/$SiO_{2}$/$VO_{2}$/substrate                                                                                     & \begin{tabular}[c]{@{}c@{}}Tuning the nanodisk size could give multiple color changes. \end{tabular}                                                             & \begin{tabular}[c]{@{}c@{}} Complex patterning of nanodisks. \\ Limited color gamut. \end{tabular} & 
 \cite{shu2018dynamic} \\ \hline
$VO_{2}$/Au(nanoparticles)/$VO_{2}$                                                                                     & \begin{tabular}[c]{@{}c@{}}By changing the size of the nanoparticles, a wide tunability for the resonance peak is possible. \end{tabular}                                                             & \begin{tabular}[c]{@{}c@{}}  High cost due to Au. \\  Limited color gamut. \end{tabular} & \cite{zhou2018modified} \\ \hline
$VO_{2}$/Au(Self - organized particles) & \begin{tabular}[c]{@{}c@{}}Self-organized Au network upon heating. Lithography-free fabrication process. \end{tabular}    & \begin{tabular}[c]{@{}c@{}} High thermal budget fabrication process. \\ High cost due to Au.   \end{tabular} & \cite{in2020self} \\ \hline
Ag/$VO_{2}$/Al & \begin{tabular}[c]{@{}c@{}}Varying the thicknesses of $VO_{2}$ and Ag produced different colour hues and brightness. \\ The color change was observed on both rigid and flexible substrates. \end{tabular}    & \begin{tabular}[c]{@{}c@{}} Multilayerd strucutre. \\ High cost due to Ag.   \end{tabular} & \cite{zhao2021flexible} \\ \hline
This work ($VO_{2}$/Al) & \begin{tabular}[c]{@{}c@{}c@{}} Simple two-layer structure. \\ No complex patterning involved. \\ Simple, low thermal budget fabrication process. \end{tabular} & \begin{tabular}[c]{@{}c@{}} Need precise control of oxidation time. \\  \end{tabular} &  \\ \hline
\end{tabular}%
}
\end{table}

\end{document}